\documentstyle[epsf,11pt]{article}
\begin{document}
\begin{center}
{\Large \bf
Numerical Study of Inelastic Scatterings
by Time-Dependent\\ Random Potentials
in Two-Dimensional Systems\\
}
\medskip\smallskip\smallskip
{\large Takeshi Nakanishi and Tomi Ohtsuki$^1$}\\
\smallskip\smallskip\smallskip
{\large \sl The Institute of Physics and Chemical Research (RIKEN)}\\
{\large\sl 2--1 Hirosawa, Wako-shi, Saitama 351--01, Japan}\\
\smallskip\smallskip
${\ }^1${\large\sl Department of Physics, Sophia University}\\
{\large\sl Kioi-cho 7--1, Chiyoda-ku, Tokyo 102, Japan}\\
\bigskip
\begin{minipage}[t]{147mm}
Diffusion of electrons in a
two-dimensional system with time-dependent random potentials
is investigated numerically.
In the absence of spin-orbit scattering, the conductivity shows
universal weak localization correction.
In the presence of it, however, the correction to the conductivity
weakly depends on the strength of disorder, and becomes vanishingly
small close to the metal-insulator transition point.
\end{minipage}
\end{center}

\bigskip
\noindent
{\large \bf 1. Introduction}
\smallskip

\noindent
Transport properties of the two-dimensional disordered electron systems have
attracted much attention in recent years.
To understand the transport properties of a random system, the
concept of quantum interference plays an important
role.
Effects of quantum interference are suppressed in the
presence of the inelastic scattering, for instance, due to
the electron-phonon interaction, due to the Coulomb interaction
of conduction electrons, or due to the motion of a single impurity atom.
The purpose of this paper is to calculate numerically the conductivity of the
two-dimensional systems with time-dependent random potentials
with and without spin-orbit scatterings, and to
demonstrate dephasing by the time-dependent potential.

The two-dimensional system with spin-orbit interactions
belongs to the symplectic universality class, and
exhibits the Anderson transition.
On the other hand, it belongs to the orthogonal universality class without
spin-orbit scatterings.
The study of the effect of dynamical potential on the conductivity
in  the critical regime  as well as the weakly localized regime
gives how the dephasing modifies the quantum interference effects
in the above two universality classes.

\bigskip
\noindent
{\large \bf 2. Model and Method}
\smallskip

\noindent
We consider the tight-binding Hamiltonian with time-dependent
potential $\varepsilon_i (t)$ on the two-dimensional square lattice:
\begin{equation}
H(t)\!=\!-\sum_{(i,j),\sigma,\sigma'}
V_{i,\sigma;j,\sigma'}C_{i,\sigma}^{\dagger}
C_{j,\sigma'}\!+\!\sum_{i,\sigma} \varepsilon_i (t)C_{i,\sigma}^{\dagger}
C_{i,\sigma},
\label{eqn:hamiltonian}
\end{equation}
with the transfer integral in the Ando model\cite{Ando 1989}
\begin{equation}
V_{i;i+\hat{x}}=\left(
\begin{array}{cc}
V_1 & V_2 \\
-V_2 & V_1
\end{array}
\right),
V_{i;i+\hat{y}}=\left(
\begin{array}{cc}
V_1 & -i V_2 \\
-i V_2 & V_1
\end{array}
\right)
\end{equation}
where $C^{\dagger}_{i,\sigma}(C_{i,\sigma})$ denotes a creation (annihilation)
operator of an electron at the site $i$ with spin $\sigma$ and $(i,j)$
are the nearest neighboring sites.
Here $\hat{x}(\hat{y})$ denotes the unit vector in the $x(y)-$direction.
All the length-scales are measured in units of the lattice constant $a_0$.
The strength of the spin-orbit interaction is characterized by the
parameter $S=V_2 /V$ with $V=\sqrt{V_1^2 + V_2^2}$.

In order to take into account the effect of moving potential,
we assume that the site-potentials take the form:
\begin{equation}
\varepsilon_{\vec{r}}(t)\!=\!\varepsilon_{\vec{r}}(0) \sqrt{2}
\cos{\left[\omega t + \theta\left(\vec{r}\right)\right]},
\label{eqn:modelpotential}
\end{equation}
where $\omega$ is the frequency.
Effects of scattering from impurities are introduced through
randomness of  site energy $\varepsilon_{\vec{r}}(0)$ and phase
$\theta(\vec{r})$ at $t\!=\!0$ distributed uniformly in the regions,
$|\varepsilon_{\vec{r}}(0)| < W/2$ and
$|\theta(\vec{r})|  <  \pi$.
We consider the adiabatic case $\omega \ll V/\hbar$, where impurities
move slower than electrons.
The method of the calculations is essentially the same as those in
Refs. \cite{Nakanishi et al 1997,kawarabayashi95} where we adopt the method
based on the
fourth-order decomposition of exponential operators\cite{suzuki}.

The quantity we observe is the second moment of the wave
packet $\langle r^2 (t) \rangle_c $ defined by
\begin{equation}
\langle r^2 (t) \rangle_c \equiv \langle |\vec{r} (t)|^2
\rangle -
\langle \vec{r}(t)\rangle^2
\end{equation}
with
\begin{equation}
\langle r^2 (t) \rangle \equiv \int {\rm d}\Omega r^{d-1} {\rm d}r
r^2 |\psi(\vec{r},t)|^2,
\end{equation}
where $\psi(\vec{r},t)$ denotes the wave function at time $t$,
and $d$ the dimensionality of the system.
If the wave function extends throughout the whole system,
the second moment is expected to grow in proportion to time $t$ as
$\langle r^2 \rangle_c\!=\!2dDt$.
Here, the coefficient $D$ denotes the diffusion coefficient.
In contrast, if the wave function is localized, it is clear that the second
moment remains finite in the limit $t \rightarrow \infty$\cite{kawarabayashi95}.
In the metallic region the Einstein relation
$\sigma\!=\!e^2 \rho (E_{\mbox{F}})D$
relates the conductivity $\sigma$ to the diffusion
constant, where $\rho (E_{\mbox{F}})$ is the density of state at Fermi
energy $E_{\mbox{F}}$.

\bigskip
\noindent
{\large \bf 3. Numerical Results}
\smallskip

\noindent
We have calculated the second moment
of the wave packet $\langle r^2 \rangle_c$
at various random potential strength.
The size of the systems are 500 by 500 for the energies
$E_{\mbox{F}}/V\!=\!-1$.
We have carried out an exact diagonalization for the 20 by 20 subsystem
at the center of the system and taken the eigenfunction of the subsystem
whose eigenvalue is closest to the given energy $E_{\mbox{F}}$
as the initial wave packet.
By this procedure we can simulate the diffusion of the wave
packet whose energy is approximately equal to $E_{\mbox{F}}$.

Fig. 1 gives an example of the calculated
second moment for $\omega\!=\!10^{-2} V/\hbar$.
The second moment is proportional to the time
in a wide range up to $2\times10^{3} \hbar/V$.
We can evaluate the diffusion constant of this system with least square fit
to these data.
In the actual simulation, the quantities $\langle r^2 \rangle_c$ are
averaged over at least four samples of random potential distribution.
The density of states $\rho (E_{\mbox{F}})$ in the Einstein relation
is evaluated from the direct diagonalization or
numerical Green's function method for $\omega\!=\!0$.
We have obtained that
$\rho (E_{\mbox{F}})a_{0}^{2}V=0.281$ and $0.251$ for $W/V=2\sqrt{2}$
and $4$, respectively in the orthogonal system $S=0$, and
$\rho (E_{\mbox{F}})a_{0}^{2}V=0.304, 0.267$ and $0.222$ for $W/V=2\sqrt{2},4$
and $4\sqrt{2}$, respectively in an example of the symplectic system $S=0.5$.
The static conductivity obtained by the present method coincides with
that by Landauer formula \cite{Nakanishi et al 1997}.

In Fig. 2 we show examples of the conductivity in
the presence (open circles) and absence (filled circles) of spin-orbit
interaction
as the function of $\ln{\tilde{\omega}}$ where $\tilde{\omega}\!=\!\omega/(V/\hbar)$.
In the absence of spin-orbit interaction the conductivity linearly
depends on $\ln{\tilde{\omega}}$ as
\begin{equation}
\sigma\!=\!A\ln{\tilde{\omega}}\ +\ \sigma(0) .
\end{equation}
The coefficients $A(\approx 0.22)$
calculated by the least square fit
to the averaged conductivity, which are shown by the dotted lines,
are almost the same among these
different disorder cases, although the conductivity
$\sigma(0)$ for $\omega\!=\!0$ is significantly different from each other.
For $S=0.5$ the conductivity decreases linearly, takes a minimum on
$\ln{\tilde{\omega}}\sim-2$, and then increases with
$\ln{\tilde{\omega}}$.
For $W/V=2\sqrt{2}$ and $4$, the absolute value of the slopes on
$\ln{\tilde{\omega}}<-2$ is almost half of orthogonal case.
The slope on $\ln{\tilde{\omega}}<-2$ represented by dashed
lines slightly decreases with the increase of the disorder $W/V$, and then
it becomes almost flat for $W/V=4\sqrt{2}$.
Note that, $W/V=4\sqrt{2}=5.66\dots$ is near the 
\newpage
\newlength{\minitwocolumn}
\setlength{\minitwocolumn}{0.5\textwidth}
\addtolength{\minitwocolumn}{-0.5\columnsep}
\noindent
\begin{minipage}[t]{\minitwocolumn}
\epsfxsize=80mm
\epsfbox{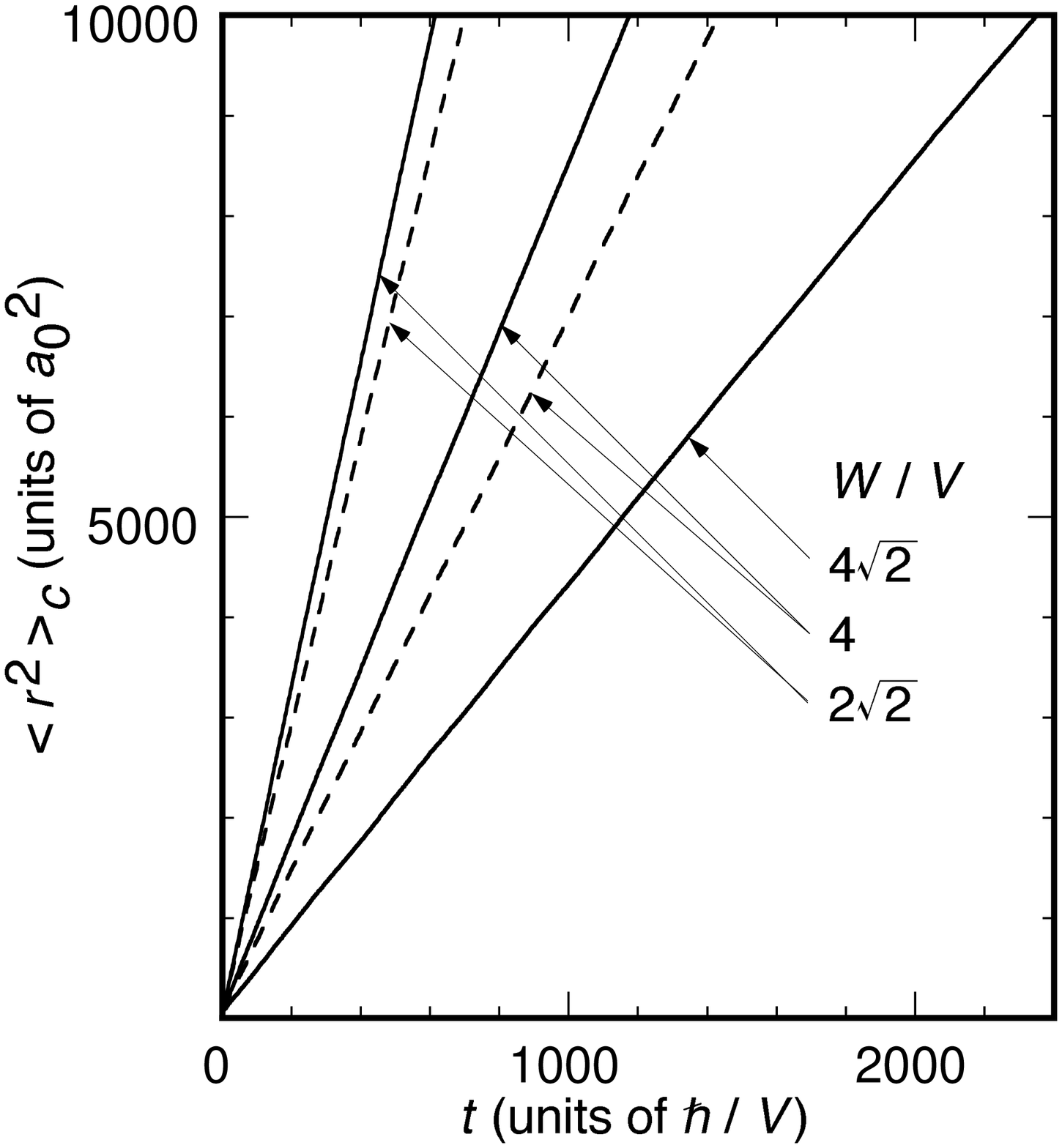}
\end{minipage}
\hspace{\columnsep}
\begin{minipage}[t]{\minitwocolumn}
\epsfxsize=80mm
\epsfbox{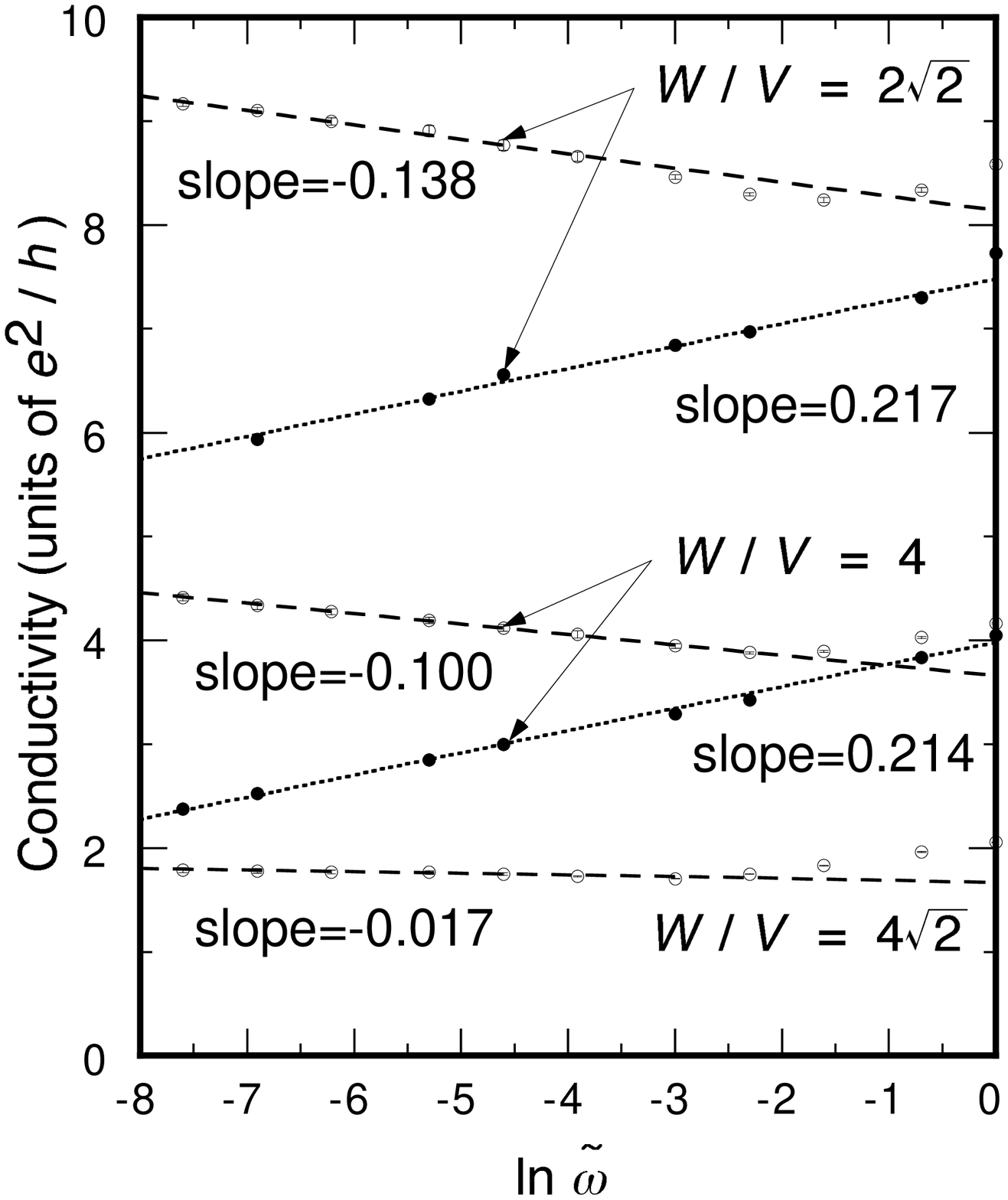}
\end{minipage}
\newline
\begin{minipage}[t]{\minitwocolumn}
Fig. 1. An example of the second moment as the function of time for $\omega\
=\ 10^{-2} V/\hbar$ and $E_{\mbox{F}}/V\!=\!-1$.
The dashed line is that in the absence of
spin-orbit interactions ($S=0$) and the solid lines in their presence
($S=0.5$).
The evaluated diffusion constants with least square fit to these data are
$D\hbar/{a_0}^2 V=3.49$ and $1.67$ for $W/V=2\sqrt{2}$ and $4$ in $S=0$, and
$D\hbar/{a_0}^2 V=4.06, 2.11$ and $1.06$ for $W/V=2\sqrt{2}, 4$ and
$4\sqrt{2}$ in $S=0.5$, respectively.
\end{minipage}
\hspace{\columnsep}
\begin{minipage}[t]{\minitwocolumn}
Fig. 2. The calculated conductivity of the two-dimensional system with time
dependent impurity potential for $W/V\!=\!2\sqrt{2},4$ and $4\sqrt{2}$.
The dotted lines indicate in orthogonal system the corrections
$\delta\sigma\!=\!0.217\frac{e^2}{h}\ln{\tilde{\omega}}$,
and $\delta\sigma\!=\!0.214\frac{e^2}{h}\ln{\tilde{\omega}}$,
 and
the dashed lines in symplectic system the corrections on
$\ln{\tilde{\omega}}<-2$
$\delta\sigma\!=\!-0.138\frac{e^2}{h}\ln{\tilde{\omega}}$,
$\delta\sigma\!=\!-0.100\frac{e^2}{h}\ln{\tilde{\omega}}$,
and $\delta\sigma\!=\!-0.017\frac{e^2}{h}\ln{\tilde{\omega}}$, respectively.
\end{minipage}\\

\bigskip
\noindent
critical point of
metal-insulator transition $W_c /V=5.74$
estimated by the finite-size scaling method for
$E_{\mbox{F}}=0$\cite{Ando 1989}.

\bigskip
\noindent
{\large \bf 4. Discussion and Conclusion}
\smallskip

\noindent
To interpret the above universal correction due to frequency $\omega$,
we recall the weak localization correction,
\begin{equation}
\delta \sigma = -\frac{e^2}{2 \pi^2 \hbar}
\ln{\frac{\tau_\phi}{\tau}},
\end{equation}
where $\tau_{\phi}$ is the dephasing time, and
$\tau$ the elastic scattering time.
The $\omega$ dependence of $\tau_{\phi}$ in the
kicked rotator problem\cite{fishman91} as well as that
in the one-dimensional  disordered system \cite{borgonovi95}
in the presence of noise
is estimated to be
\begin{equation}
\tau_{\phi} \sim \omega^{-2/3}.
\end{equation}
Since the qualitative argument leading to this dependence is
independent of the space dimension,
the 2D weak localization correction $\delta\sigma$~\cite{bergmann84,HLN}
is estimated to be
\begin{equation}
\delta \sigma =
 \frac{2}{3\pi}\frac{e^2}{h}\ln{\frac{\omega}{\omega_0}} .
\label{eqn:weaklocalization}
\end{equation}
The pre-factor $2/3\pi\!=\!0.212\dots$ of the $\ln{\omega}$ term
agrees with the numerical calculation
shown in Fig. 2,
which is universal and independent of the potential strength $W$.
Note that this $\omega$-dependence is also observed in
the case of low energy phonon scattering \cite{note}.

The results with the spin-orbit interaction is consistent with the weak
localization correction too.
For $\tau_{\phi}\ll\tau_{SO}$, where $\tau_{SO}$ is the dephasing time due
to spin-orbit interaction, the conductivity increases with $\omega$ as
in the orthogonal case, while
the correction to the conductivity is half with opposite sign for
$\tau_{\phi}\gg\tau_{SO}$.

In conclusion, we have analyzed the weak localization effect
in the two-dimensional system with time-dependent random potentials
using the equation-of-motion method.
It has been shown numerically that the weak localization correction
to the conductivity due to the fluctuating potentials
takes universal value independent of the random potential strength $W$
in orthogonal system.
The symplectic system is also investigated and it is demonstrated that the
conductivity becomes almost independent of the frequency near the critical point of
metal-insulator transition.
Such behavior is observed in the recent experiments with high mobility
Si MOS \cite{Kravchenko et al 1995}, though the mechanism leading
to the metal-insulator transition is different.
Our results will open a new way to incorporate the dephasing mechanism
in numerical simulations.

\bigskip
\noindent
{\large \bf Acknowledgments}
\smallskip

\noindent
Numerical calculations were performed on FACOM VPP500 in Supercomputer
Center at RIKEN and ISSP.


\end{document}